\documentclass{article}
\usepackage{spconf,amsmath,graphicx}
\usepackage{graphicx,algorithm,algorithmicx,bm,amsmath,amssymb,rotating,graphicx,cite,amsthm,calc,color,epstopdf,xcolor,mathrsfs,dsfont}
\usepackage{lettrine}
\usepackage{cite}
 \usepackage{amsfonts}
\usepackage{longtable}
\usepackage{float}
\usepackage{subcaption}
\usepackage{flushend}

\def\bD{\mathbb{D}}
\ninept
\title{Leveraging Foundation Models for Efficient Federated Learning in Resource-Restricted Edge Networks}
\name{S. Kawa Atapour$^\dag$, S. Jamal SeyedMohammadi$^\dag$, S. Mohammad Sheikholeslami$^\ddag$,\vspace{-.14in}} 
\address{\textit{Jamshid Abouei $^{\dag\dag}$, Konstantinos N. Plataniotis$^\ddag$, and Arash Mohammadi$^\dag$\thanks{This work was partially supported by the Natural Sciences and Engineering Research Council (NSERC) of Canada through the NSERC Discovery Grant RGPIN-2023-05654}}\\\\
$~^\dag$ Intelligent Signal \& Information Processing (I-SIP) Lab, Concodia University, Canadaa \\$~^\ddag$ Edward S. Rogers Sr. Department of Electrical and Computer Engineering, University of Toronto \\$~^{\dag\dag}$ Department of Electrical Engineering, Yazd University, Iran}
\begin{document}
\maketitle
\begin{abstract}
%Although Federated Learning (FL) supports data privacy, its performance is limited over edge networks due to high resource consumption and insufficient data of Internet of Things (IoT) devices. 
Recently pre-trained Foundation Models (FMs) have been combined with Federated Learning (FL) to improve training of downstream tasks while preserving privacy. 
However, deploying FMs over edge networks with resource-constrained Internet of Things (IoT) devices is under-explored. 
This paper proposes a novel framework, namely, Federated Distilling knowledge to Prompt (FedD2P), for leveraging the robust representation abilities of a vision-language FM without deploying it locally on edge devices. 
This framework distills the aggregated knowledge of IoT devices to a prompt generator to efficiently adapt the frozen FM for downstream tasks.
%Subsequently, the general knowledge of the FM is utilized for IoT devices to enhance their generalization ability.
To eliminate the dependency on a public dataset, our framework leverages per-class local knowledge from IoT devices and linguistic descriptions of classes to train the prompt generator.
Our experiments on diverse image classification datasets CIFAR, OxfordPets, SVHN, EuroSAT, and DTD show that FedD2P outperforms the baselines in terms of model performance.
\end{abstract}
\begin{keywords}
Federated learning, foundation models, distilling knowledge, prompt-tuning.
\end{keywords}
%
%OOOOOOOOOOOOOOOOOOOOOOOOOOOOOOOOOOOOOOOOOOOOOOOOOOOOOOOO
\section{Introduction} \label{sec:intro}
%OOOOOOOOOOOOOOOOOOOOOOOOOOOOOOOOOOOOOOOOOOOOOOOOOOOOOOOO
Traditional centralized learning over distributed Internet of Things (IoT) networks~\cite{IoT_network} fails to reach the expected performance mainly due to distribution of data over resource-constrained devices, limited communication resources, and privacy concerns. In this context, Federated Learning (FL)~\cite{FL_survey_2024} has emerged as a fruitful alternative to the centralized approach by facilitating collaborative and privacy-preserving knowledge exchange among distributed IoT devices. This is achieved through repetitive communications with a coordinating server (fusion centre), which has higher computation power, and is responsible for aggregation of the distributed knowledge~\cite{constrained_IoT}. For applications that involve training of Deep Learning (DL) models from scratch, performance of conventional FL frameworks~\cite{FedBN, from_scratch, mofleur} can significantly degrade especially in resource-limited IoT networks.  Such degradation is due to the fact that training on insufficient data distributed over edge devices requires several computation/communication rounds resulting in excessive overhead and latency. This becomes particularly significant for complex tasks under statistical and system heterogeneity. %posing large delays in recent FL schemes.

\vspace{.05in}
\noindent
\textbf{\textit{Literature Review:}} Recently to speed up learning downstream tasks, Foundation Models (FM)~\cite{FM}, which are, typically, large DL models trained on general-purpose datasets, have been utilized as the backbone of local models in FL~\cite{FM}. By leveraging few-shot capabilities of FMs in FL scenarios, clients are not required to train their models from scratch, significantly reducing the communication/computation overhead~\cite{promptfl}. Additionally, the pre-trained knowledge incorporated in FMs can effectively mitigate the data scarcity issue~\cite{FedTPG}. To adapt FMs to downstream tasks, fine-tuning methods~\cite{partial_fine-tuning, Adapter, coop} have been employed.  Among these methods, prompt-tuning is more adaptable to FL in edge environments, as it requires less computational resource and storage space, and outperforms its alternatives when dealing with limited data~\cite{elephent}. In prompt-tuning, a prompt is given to a pre-trained FM to generate specific responses corresponding to the downstream task without the need for additional training or gradient updates on the FM. 

Consequently, there has been a recent surge of interest~\cite{promptfl, Fedprompt, pFedPG,pFedPrompt, FedTPG, fedclip} to adapt pre-trained FMs to downstream tasks using prompt tuning in collaborative frameworks. A drawback of these prior works is overlooking availability constrained resources at the client side, making them impractical for distributed IoT networks. There are few recent attempts to address this shortcoming, for instance, FedHPL~\cite{FedHPL} allows clients to download resource-appropriate versions of FMs from the server. Furthermore, FedHPL targets handling the heterogeneity of local models by employing the knowledge distillation technique, where logits, instead of prompts, are shared with the server. While FedHPL reduces the computation costs associated with fine-tuning of local FMs, it still assumes that clients possess sufficient storage for deploying and prompt tuning of local FMs. To further tackle resource constraints on devices, FedMKT~\cite{FedMKT} proposed to deploy an Large Language Model (LLM), such as LLaMa-2 with 7 billion parameters~\cite{llama2}, on the server while pre-trained small language models (e.g., GPT-2 with 1.5 billion parameters~\cite{survey_on_GPT2}) are placed on the local devices. Although FedMKT liberates resource-constrained clients from performing local prompt tuning of FMs, the small pre-trained FMs are still relatively large, requiring more resources than IoT devices can, typically, provide. The paper aims to address this gap.

\vspace{.05in}
\noindent
\textbf{\textit{Contributions:}} To address the above mentioned issue, we introduce the Federated Distilling knowledge to Prompt (FedD2P) framework, which strategically places the FM exclusively on the server, eliminating the need for FM deployment on IoT devices. More specifically, the distributed knowledge from the lightweight local models of IoT devices is distilled into a prompt generator module, facilitating adaptation of a vision-language FM with the downstream task.
Subsequently, the robust knowledge of the FM is employed to improve the generalization capabilities of IoT devices. By centralizing the FM on the server, IoT devices can deploy smaller models based on their available resources, enabling FedD2P to also effectively handle model heterogeneity.  In summary, the paper makes the following key contributions:
\begin{itemize} 
\item[\textbf{[C1]}] Design of the FedD2P framework wherein distributed, resource-constrained IoT devices leverage the robust representational knowledge of a vision-language FM located at the server.
\item[\textbf{[C2]}] Introduction of a novel data-free mutual Knowledge Distillation (KD) framework based on per-class knowledge transfer of IoT devices and the sever side's FM. In this regard, a linguistic assistance prompt generator is designed, which is fine-tuned using the distributed per-class knowledge of IoT devices and the linguistic descriptions of the classes.
\end{itemize}

The rest of the paper is organized as follows: Section~\ref{sec:format} provides background information required for presentation of the proposed FL approach. The FedD2P framework is then introduced in Section~\ref{sec:FedD2P}. The simulation results are presented in ~\ref{simulations}, and, finally, Section~\ref{conclusion} concludes the paper.

%OOOOOOOOOOOOOOOOOOOOOOOOOOOOOOOOOOOOOOOOOOOOOOOOOOOOOOOO
\section{Background and System Model} \label{sec:format}
%OOOOOOOOOOOOOOOOOOOOOOOOOOOOOOOOOOOOOOOOOOOOOOOOOOOOOOOO
%%%%%%%%%%%%%%%%%%%%%%%%%%%%%%%%%%%%%%%%%%%%%%%%%%%%%
\setlength{\textfloatsep}{0pt}
\begin{figure*}[t!]
\centering
\includegraphics[width=0.7\textwidth]{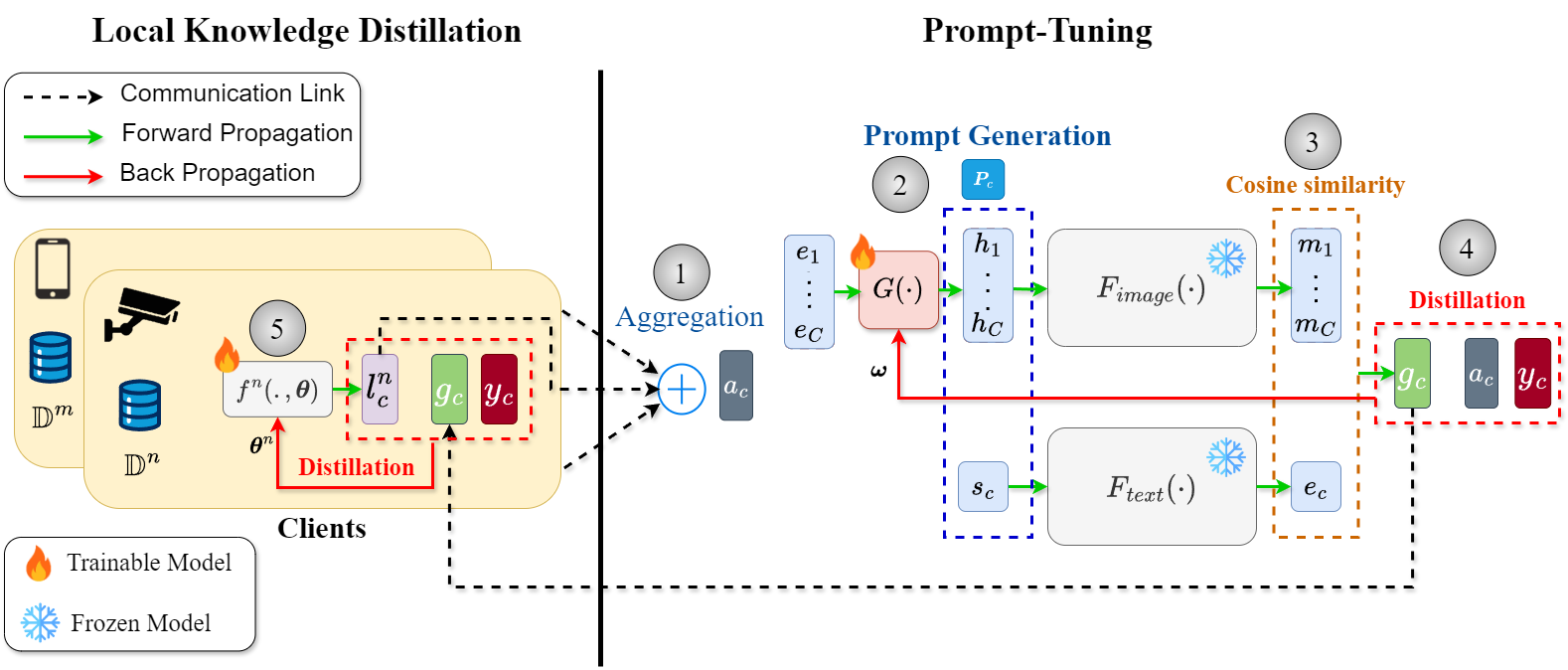}
\caption{\footnotesize The proposed FedD2P framework. In 1) the per-class local knowledge of IoT devices, denoted as $\boldsymbol{l}^n_c, $ for ($1 \leq n \leq N$) are aggregated at the server, resulting in the per-class global knowledge $\boldsymbol{a}_c$. In 2) the LA prompt generator poduces per-class prompts  $[\boldsymbol{h}_c]_{c=1}^C$ using the semantic representation of classes $[\boldsymbol{e}_c]_{c=1}^C$. Subsequently, the image and text encoders generate semantic features for their respective prompts, i.e.,  $[\boldsymbol{m}_c = F_{image}(\boldsymbol{h}_c)]_{c=1}^C$ , $\boldsymbol{e}_c = F_{text}(\boldsymbol{s}_c)$ respectively.  The per-class global knowledge $\boldsymbol{g}_c$ is subsequently determined by calculating the cosine similarity between these semantic features. In 4) the per-class aggregated knowledge $\boldsymbol{g}_c$ and ground-truth output $\boldsymbol{y}_c$ are used to tune the LA generator, while the backbone FM remains freezed. Finally, in 5), the global knowledge is transmitted to IoT devices to facilitate local knowledge distillation.}
\label{fig:FedD2P}
\end{figure*}
%%%%%%%%%%%%%%%%%%%%%%%%%%%%%%%%%%%%%%%%%%%%%%%%%%%%%

In this section, first, we briefly present the required background on KD technique and prompt-tuning for FL. Then, we present the system setup and formulate the problem of FL over a resource-limited IoT network. 

%==============================================================
\subsection{Knowledge Distillation (KD)}
%==============================================================
Generally speaking, KD refers to the method of transferring knowledge from one or multiple teacher models to a student model~\cite{KD}. Specifically, the softened output of the teacher model on a public dataset, along with the ground-truth output, is used to train the student model. In this context, a KD loss function is employed in addition to the supervised learning loss to minimize the discrepancy between the soft labels of the teacher model and the predictions made by the student model. In KD assisted FL scenarios, depending on the method, both clients and the server can assume the roles of either the teacher or the student, i.e., mutual KD~\cite{fedd2s}. In such scenarios, knowledge from clients is shared with the server to create a global knowledge base. This knowledge is then distilled back to the clients to enhance their performance, allowing for the transfer of knowledge from other clients to each individual client. Typically, KD methods~\cite{knfu} utilize a public dataset shared among all entities (also referred to as the transfer set) to align the extracted knowledge of local models and the server's one. This assumption of a publicly available dataset, however, is unrealistic in practice, as it can be accessed by third parties and raise privacy concerns. In this paper, KD is employed in a data-free fashion (without reliance of a transfer set), and to establish a framework for knowledge exchange between local models and the FM of the server.

%==============================================================
\subsection{Prompt Tuning}
%==============================================================
Fine-tuning FMs for downstream tasks has shifted the dominant paradigm of machine learning from ``training from scratch'' to the ``pretrain-then-finetune'' framework. Fully fine tuning of a large FM, however, involves updating all its parameters, which increases the risk of overfitting. This challenge has driven the development of partial fine-tuning methods~\cite{partial_fine-tuning}. With the advent of LLMs, a novel capability known as prompting~\cite{prompting_survey} has been introduced for such models.  This technique involves prepending learnable parameters to the embedding space of the input—whether it be the embedding space of tokens in LLMs or the embedding space of image patches in Transformer-based vision models. This approach provides pre-trained FMs with hints about downstream tasks while keeping their parameters frozen. For further details please refer to Reference~\cite{details_of_prompting}.

%AC: As we have space and time, for completeness, possibly you can add some basic formulation for prompt tuning here.

%==============================================================
\subsection {Federated Learning over Edge}
%==============================================================
We consider an edge environment consisting of $N$ IoT devices, denoted by $\mathbb{U}=\{u_1, \cdot\cdot\cdot,u_N\}$, which are coordinated by an edge server.  Each IoT device $u_n$ for ($1 \leq n \leq N$), aims to perform a $C$-class classification task with the assistance of an FM located at the edge server.
Each IoT device $u_n\in\mathbb{U}$ possesses a local dataset, represented as $\bD^n$.
Local datasets are distributed heterogeneously among IoT devices and collectively form the entire dataset $\bD=\{\bD^1, \cdot\cdot\cdot,\bD^N\}$. Each IoT device, based on its available storage and computation resources, employs a lightweight local model, denoted by  $f^n(\cdot; \boldsymbol\theta^n)$ parameterized by $\boldsymbol\theta^n$.
This enables the framework to handle model heterogeneity. To develop a FL framework that trains local models with the support of an FM at the server, we solve the following personalized FL problem
\begin{equation}
\begin{array}{rrclcl}
	\operatorname*{argmin}_{\{\boldsymbol\theta^1,\cdot\cdot\cdot,\boldsymbol\theta^n\}} \mathbb{E}_{\bD^n\in\bD}\{ \mathcal{J}^n(\boldsymbol\theta^n, \mathcal{D}^n) \},
\end{array}
\end{equation}
where $\mathcal{J}^n(\cdot)$ is the loss function of local model $f^n(\cdot, \boldsymbol \theta^n)$ over local dataset $\bD^n$.

%OOOOOOOOOOOOOOOOOOOOOOOOOOOOOOOOOOOOOOOOOOOOOOOOOOOOOOOO
\section{Federated Distilling Knowledge To Prompt (F\lowercase{ed}D2P)} \label{sec:FedD2P}
%OOOOOOOOOOOOOOOOOOOOOOOOOOOOOOOOOOOOOOOOOOOOOOOOOOOOOOOO

The proposed FedD2P framework operates through a repetitive workflow, where in each communication round, the local knowledge of IoT devices is shared with the server to construct the aggregated knowledge in form of soft labels. This knowledge is then utilized to fine-tune the FM for the downstream task. Distilling the aggregated knowledge of IoT devices can more effectively instruct the FM towards the downstream task. The fine-tuned FM then generates the global knowledge, which is subsequently transmitted back to the IoT devices to assist them in local training. 

To establish a data-free KD framework, we adopt the  per-class knowledge sharing approach instead of per-sample knowledge transfer. This approach aligns the extracted knowledge of local models and the FM at the class level, restricting the amount of local knowledge shared by the IoT devices. To compensate this knowledge scarcity, we utilize the information in linguistic description of classes.  The rich semantic representation of linguistic content from vision-language FMs can be employed for this purpose. Specifically, we propose a Linguistic Assistance (LA) prompt generator to facilitate this process. Fig.~\ref{fig:FedD2P} provides the flow of knowledge of the proposed FedD2P framework and the LA prompt generator. Next, we provide further details on different components of the proposed FedD2P framework.

%==============================================================
\subsection{The Flow of knowledge}
%==============================================================
The proposed FedD2P framework includes an initialization stage followed by iterating four steps, summarized as follows:

\vspace{.05in}
\noindent 
\textbf{\textit{(S0) Initialization:}} To initiate the FL process, firstly, each IoT device trains its local model on the local dataset, $\bD^n$.

%AC: Define \bm{X}_c^n either here or where you formulate the problem.
\vspace{.05in}
\noindent 
\textbf{\textit{(S1) Knowledge Aggregation:}} The local knowledge of IoT device $u_n$ for class $c$ is computed by averaging soft labels generated by $f^n(\cdot;\boldsymbol \theta^n)$ on $\boldsymbol{X}_c^n$, as 
\begin{eqnarray}
\boldsymbol{l}^n_{c}=\frac{1}{|\boldsymbol{X}^n_c|}\sum_{x\in \boldsymbol{X}_c^n} \sigma_{\tau}\big(f^n(x,\boldsymbol\theta^n))\big), 
\end{eqnarray}
where $\boldsymbol{X}_c^n$ denotes input samples of dataset $\bD^n$ that belong to class $c$, and $\sigma_{\tau}(\cdot)$ represents the softmax function with the temperature parameter $\tau$.
These per-class local knowledge representations are then averaged at the server to form the per-class aggregated knowledge, computed as follows $\boldsymbol{a}_c = \sum_{n=1}^{N}\frac{|\boldsymbol{X}^n_c|}{|\boldsymbol{X}_c|} \boldsymbol{l}^n_c$,
where, $\boldsymbol{X}_c$ denotes the set of all input samples from the dataset $\bD$ associated with class $c$.

\vspace{.05in}
\noindent 
\textbf{\textit{(S2) Fine-tuning the FM:}}  The aggregated knowledge is subsequently utilized to fine-tune the FM for the downstream task will be described later in Subsection \ref{sec:LAproGen}. Following this fine-tuning, the per-class global knowledge of the FM is computed as $\boldsymbol{g}_c = \sigma_{\tau}\big(F(\boldsymbol{P} _c, \boldsymbol{\phi})\big)$, where $F(\cdot; \boldsymbol{\phi})$ denotes the FM with parameters $\boldsymbol{\phi}$, and $\boldsymbol{P}_c$ represents the class-specific prompt.

\vspace{.05in}
\noindent 
\textbf{\textit{(S3) Local Knowledge Distillation:}}  Per-class global knowledge, $\boldsymbol{g}_c$, is then transmitted to each IoT device $u_n$, for ($1 \leq n \leq N$), to perform local knowledge distillation as follows
\begin{equation}
 	 \operatorname*{argmin}_{\boldsymbol \theta^n} \sum_{c=1}^{C}\sum_{x\in \boldsymbol{X}^n_c} \mathcal{L}_{CE}\big(f^n(x,\boldsymbol \theta^n), y\big) + \mathcal{L}_{KL}\big(\sigma_{\tau}(f^n(x,\boldsymbol \theta^n)), \boldsymbol{g}_c\big),
\end{equation}
where $\mathcal{L}_ {CE}(\cdot)$, and $\mathcal{L}_ {KL}(\cdot)$ are per-sample cross-entropy and Kullback-Leibler loss functions, respectively.
Here,  $y$ represents the ground-truth corresponding to the input sample $x$.

%==============================================================
\subsection{Linguistic Assistance Prompt Generation}\label{sec:LAproGen}
%==============================================================
To construct the LA prompt generator and without loss of generality, we assume that the linguistic description of classes within the downstream task are available at the server and are represented by $\{s_1, \ldots ,s_C\}$.
Their corresponding semantic representation vectors are extracted by the text encoder of the vision-language FM as $\boldsymbol{e}_c = F_{text}(s_c)  \in \mathbb{R}^{d}$, where $F_{text}(\cdot)$ denotes the text encoder of the FM, and $d$ represents its embedding dimension. The prompt generator $G(\cdot;\boldsymbol \omega)$, then uses these semantic representation of linguistic description of classes to generate class-specific prompts. To account for the correlation among the semantic representations of classes, we employ a multi-head self-attention mechanism to construct the LA prompt generator. The prompt for class $c$ is, therefore, calculated as
\begin{equation}
 	\boldsymbol{h}_c = \text{Softmax}(\frac{\boldsymbol{q}_c\boldsymbol{\mathcal{K}}}{\sqrt{d}})\boldsymbol{\mathcal{V}}\boldsymbol{\mathcal{W}}_{h},
\end{equation}
where $\boldsymbol{q}_c= \boldsymbol{e}_c \boldsymbol{\mathcal{W}}_{q}, \boldsymbol{\mathcal{K}}=\boldsymbol{E}\boldsymbol{\mathcal{W}}_{k}$, and $\boldsymbol{\mathcal{V}}=\boldsymbol{E}\boldsymbol{\mathcal{W}}_{v}$ represent the query vector, and the key and value weight matrices, respectively. Here, $ \boldsymbol{E} = [\boldsymbol{e}_c]_{c=1}^C$, and $\mathcal{W}_{h}$ denotes the parameters of the head layer.
Accordingly, the class-specific prompts is defined as $\boldsymbol P_c=\{s_c, \boldsymbol{h}_1,\cdot\cdot\cdot,\boldsymbol{h}_C\}$, which is forwarded to the vison-language FM to generate the global knowledge as follows
\begin{equation}
	\boldsymbol{g}_c = \frac{\text{exp}\big(\text{cos}(   F_{image}(\boldsymbol{h}_c), F_{text}(s_c) )\big)}{ \sum_{c=1}^{C} \text{exp}\big(\text{cos}(   F_{image}(\boldsymbol{h}_c), F_{text}(s_c) )\big) },
\end{equation}
where $F_{image}(\cdot)$ denotes the image encoder of the FM.

To distill the per-class aggregated knowledge $\boldsymbol{a}_c$ into the LA prompt generator, the prompt-tuning is performed as follows
\begin{equation}
 \operatorname*{argmin}_{\boldsymbol \omega} \sum_{c=1}^{C} \mathcal{L}_c (\boldsymbol{g}_c, y_c) + \mathcal{L}_{k}(\boldsymbol{g}_c, \boldsymbol{a}_c),
\end{equation}
 where $y_c$ represents the ground-truth corresponding to the class $c$. This completes presentation of the proposed FedD2P framework, next, evaluation experiments are presented.

%OOOOOOOOOOOOOOOOOOOOOOOOOOOOOOOOOOOOOOOOOOOOOOOOOOOOOOOO
\vspace{-.15in}
\section{Simulation Results}\label{simulations}
\vspace{-.1in}
%OOOOOOOOOOOOOOOOOOOOOOOOOOOOOOOOOOOOOOOOOOOOOOOOOOOOOOOO
%%%%%%%%%%%%%%%%%%%%%%%%%%%%%%%%%%%%%%%%%%%%%%%%%%%%%%%%%%%
\begin{table*}[h]
\caption{\footnotesize Comparison of Average Test Accuracy (\%) for Five Image Classification Tasks Under Homogeneous and Heterogeneous Statistical Distributions. \textbf{Bold} means the best. \label{table}}
\centering
\begin{tabular}{c| c c| c c| c c| c c |c c}
\hline
Baseline      & \multicolumn{2}{ c}{CIFAR10} & \multicolumn{2}{c}{ SVHN} &  \multicolumn{2}{c}{ OxfordPets}& \multicolumn{2}{c}{ EuroSAT} & \multicolumn{2}{c}{DTD}\\
\hline
                & $\alpha=0.1$ & $\alpha=10$ & $\alpha=0.1$ & $\alpha=10$ &  $\alpha=0.1$ & $\alpha=10$ &  $\alpha=0.1$ & $\alpha=10$ &  $\alpha=0.1$ & $\alpha=10$\\
\hline
\textbf{B1}   &   66.16 &  67.25  & 81.57   & 88.31   & 61.37 &   64.11 & 77.52 & 81.27  & 63.47  & 70.32    \\
\hline
\textbf{B2}   &   68.35 &  70.39  & 85.10   & 88.24   & 65.34 &   67.48 & 77.60 & 80.89  & 67.90.   & 72.50    \\
\hline
\textbf{B3}   &   74.01 &  73.21 &  90.81   & 93.66   & 74.49 &   78.10 & 83.71 & 84.26  & 70.77  & 74.29  \\
\hline
FedD2P        &   \textbf{75.95} &  \textbf{77.36} &  \textbf{92.24}   & \textbf{94.36}   & \textbf{77.01} &   \textbf{78.92} & \textbf{86.35} & \textbf{88.03} & \textbf{72.18}   & \textbf{75.37}    \\
\hline
\end{tabular}
\end{table*}
%%%%%%%%%%%%%%%%%%%%%%%%%%%%%%%%%%%%%%%%%%%%%%%%%%%%%%%%%%%
In this section, we present different experiments conducted to evaluate performance of the proposed framework. We consider a distributed IoT network consisting of $10$ devices with data and model heterogeneity, collectively performing FL over $20$ communication rounds. The FL process is orchestrated by an edge server with high computation power employing a Contrastive Language-Image Pre-Training (CLIP) model as the vision-language FM. For statistical data heterogeneity, we employ the Dirichlet distribution $D(\alpha)$ with two different settings: $\alpha=10$ for homogeneous, and $\alpha=0.1$ for heterogeneous setting. We deploy Convolutional Neural Networks (CNNs) as local models, consisting of Visual Geometry Group (VGG) blocks. For model heterogeneity, we randomly select a CNN model with $2$, $3$, or $4$ VGG blocks for each device.  The size of the each local dataset is set to $4,000$ samples. The number of local epochs and batch size are set to $10$ and $128$, respectively. In each round, the LA prompt generator is trained for $100$ rounds. We set the temperature parameter as $\tau_1 = 10$ and $\tau_2 = 0.1$ for softening local and global outputs, respectively.
The selection of hyperparameters, specifically the number of rounds and epochs, is determined empirically. In contrast, the selection of temperature parameters is elaborated upon in Section \ref{sec: temperature parameters}.

%AC: add a line here on how these parameters are computed, e.g., emperically or based on reference~\cite{}, and if empirically give some insights on how to perform hyper-parameter selection.

\vspace{.05in}
\noindent 
\textbf{\textit{Datasets:}} We employ the CIFAR10~\cite{cifar10_dataset} and SVHN~\cite{svhn_dataset} datasets for general object classification, while the OxfordPets dataset~\cite{oxfordpets_dataset} is utilized for fine-grained classification tasks. Additionally, the EuroSAT~\cite{eurosat} and DTD~\cite{dtd_dataset} datasets are used for specialized tasks involving satellite imagery and texture recognition, respectively. For the CIFAR10 and OxfordPets datasets, the linguistic description of class $c$ is ``a photo of [c]''. For the EuroSAT and DTD datasets, the descriptions are ``[c] texture'' and ``a centered satellite photo of [c]'', respectively. For the SVHN dataset, we use ``a photo of digit [c]''. Here [c] denotes the name of class $c$ in natural language.

\vspace{.05in}
\noindent 
\textbf{\textit{Baselines:}}The comparison of the proposed FedD2P framework with baselines ~\cite{FedMKT} and ~\cite{FedHPL}, which utilize significantly more powerful local models, is not fair within our settings. This is due to the limited computational resources of edge devices, which prevent the possibility of maintaining a FM locally.
Therefore, we compare FedD2P against three baselines: \textbf{B1)} Each client trains its model solely on its local dataset without participating in the FL process, \textbf{B2)} The aggregated knowledge, rather than the global model, is transmitted back to clients, and \textbf{B3)}, Only the ground-truth output $y_c$ is used to fine-tune the LA prompt generator, while the per-class aggregated knowledge $\boldsymbol{a}_c$ is not distilled into it.
%AC: add a line why we cannot have a fair comparison with other, references can not be implemented in our settings given local computational resources of local nodes eliminating possibility of maintaining a FM locally. 

%==============================================================
\vspace{-.15in}
\subsection{Evaluation Under Different Statistical Heterogeneity}
\vspace{-.05in}
%==============================================================
Table~\ref{table} presents the performance comparisons of FedD2P with the baselines across two statistical distribution settings. The results indicate that leveraging an FM at the server is more effective than relying solely on the aggregated knowledge from edge devices.
We attribute this consistently superior performance to the robust representations provided by the CLIP model and the effective distillation of global knowledge to IoT devices.
It can also be inferred that distilling the distributed knowledge of IoT devices to the LA prompt generator fine-tunes the FM for the downstream task more effectively.
The most significant performance gain compared with other baselines is observed for the OxfordPets dataset.
This is likely due to the dataset’s challenging nature for training a local CNN model from scratch, whereas according to ~\cite{coop} the few-shot accuracy of the CLIP model on this dataset is superior compared with other datasets. 
%AC: please expand on this more intuitively

%==============================================================
\vspace{-.15in}
\subsection{Evaluation Under Different Temperature Parameter}\label{sec: temperature parameters}
\vspace{-.05in}
%==============================================================
The effectiveness of the mutual KD framework is significantly determined by the entropy of soft labels. 
Low entropy for the local models and high entropy for the CLIP model necessitate adjusting the temperature parameters when computing local and global soft labels. 
Fig.~\ref{fig:temp} illustrates the performance of the FedD2S framework over CIFAR10 dataset and across different temperature parameters.
As shown, the most effective performance is achieved by decreasing the entropy of global soft labels and increasing the entropy of local ones. 
This can be attributed to the high generalization ability of the robust CLIP model and the low generalization capability of simple local models.

%==============================================================
\vspace{-.15in}
\subsection{Effectiveness of LA prompt generator}
\begin{figure}[t!]
    \centering
    \begin{subfigure}[b]{0.23\textwidth}
        \centering
        \includegraphics[width=\textwidth]{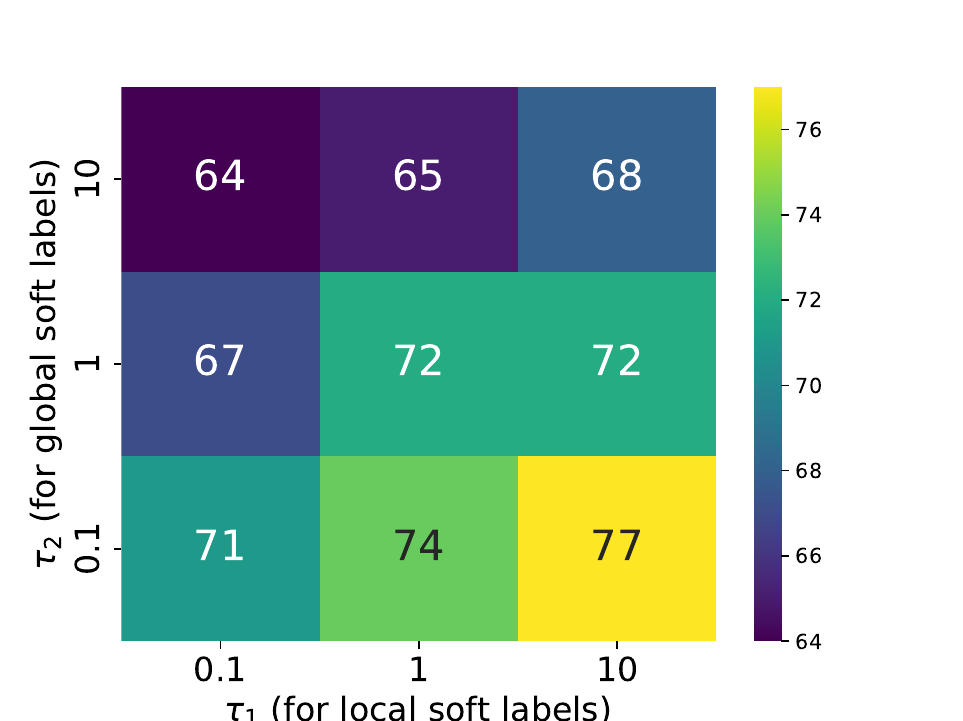}
        \caption{Effect of the temperature parameter}
        \label{fig:temp}
    \end{subfigure}
    \hfill
    \begin{subfigure}[b]{0.23\textwidth}
        \centering
        \includegraphics[width=\textwidth]{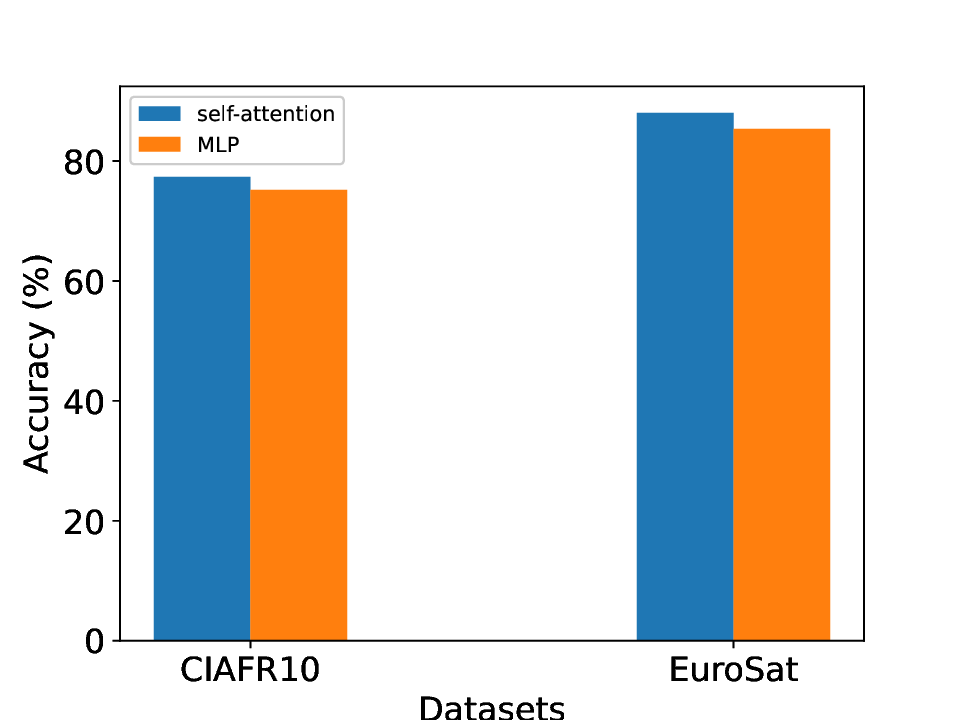}
        \caption{Effectiveness of the self-atention mechanism}
        \label{fig:la}
    \end{subfigure}
    \caption{\footnotesize (a) Sensitivity of the FedD2P framework to the temperature parameter. (b) Effectiveness of the multi-head self-attention mechanism in the LA prompt generator.}
    \label{fig:com}
\end{figure}

\vspace{-.05in}
%==============================================================
For comparison, we employ a Multi-Layer Perceptron (MLP)-based prompt generator, wherein each semantic representation vector, $\boldsymbol{e}_c$, is directly mapped to $\boldsymbol{h}_c$.
Fig. \ref{fig:la} illustrates the impact of the self-attention mechanism within the LA prompt generator module on the CIFAR10 and EuroSat datasets.
The superior performance of the multi-head self-attention mechanism is attributed to its ability to consider the relationships among different semantic representations of classes.
This capability enables the generation of effective prompts that extract soft labels with sufficient generalization content, thereby assisting local models.

%%%%%%%%%%%%%%%%%%%%%%%%%%%%%%%%%%%%%%%%%%%%%%%%%%%%%
%%%%%%%%%%%%%%%%%%%%%%%%%%%%%%%%%%%%%%%%%%%%%%%%%%%%%
%OOOOOOOOOOOOOOOOOOOOOOOOOOOOOOOOOOOOOOOOOOOOOOOOOOOOOOOO
\vspace{-.15in}
\section{Conclusion}\label{conclusion}
\vspace{-.1in}
%OOOOOOOOOOOOOOOOOOOOOOOOOOOOOOOOOOOOOOOOOOOOOOOOOOOOOOOO
Our proposed FedD2P framework successfully leverages the robust representation abilities of vision-language FMs without deploying them locally on resource-constrained edge devices. 
By distilling aggregated knowledge from IoT devices to a prompt generator, FedD2P enhances the efficiency and performance of local models in diverse image classification tasks. 
Extensive simulations demonstrate that FedD2P not only achieves competitive performance compared to traditional baselines but also significantly improves local resource efficiency, making it a promising solution for federated learning in edge networks.

\label{sec:typestyle}
\footnotesize
\bibliographystyle{IEEEtran}
\bibliography{strings}

\end{document}